# Securing Speech in GSM Networks using DES with Random Permutation and Inversion Algorithm

Khaled Merit[1] and Abdelazziz Ouamri[2]

[1] National Institute of Telecommunications and Information and communication technologies, INT&TIC Oran, Algeria
`merit1984@gmail.com`

[2] Signals and Images Laboratory, University of Sciences and Technology of Oran USTO Oran, Algeria
`ouamri@univ-usto.dz`

**ABSTRACT**

*Global System for Mobile Communications (GSM) is one of the most commonly used cellular technologies in the world. One of the objectives in mobile communication systems is the security of the exchanged data. GSM employs many cryptographic algorithms for security like A5/1, A5/2 and A5/3. Even so, these algorithms do not provide sufficient level of security for protecting the confidentiality of GSM. Therefore, it is desirable to increase security by additional encryption methods. This paper presents a voice encryption method called: "DES with Random permutation and Inversion", based on current voice channel, which overcomes data channel's insufficiencies and solves the problem of penetrating the RPE-LTP vocoder by the encrypted voice. The proposed method fulfils an end-to-end secured communication in the GSM; insure a good compatibility to all GSM networks, and easy implementation without any modification in these systems.*

## 1. INTRODUCTION

Security presents a very important axis in wireless communication systems. This is obviously because of the ubiquitous wireless medium's nature that makes it more susceptible to attacks. Any eavesdropper can get over to whatever is being sent over the network through the wireless medium. In addition, the presence of communication does not uniquely identify the originator. Besides this, any eavesdropping or tapping cannot even be detected in a medium as ubiquitous as the wireless medium which makes the latter situation even worse. Hence, security plays a fundamental task for the successful operation of a mobile communication system.

To secure data in GSM, encryptions and mechanisms to grant it are obligatory. In this paper, a new approach has been proposed which includes extra encryption DES with random permutation and inversion algorithm. GSM employs stream ciphers for encryption which requires the data to be in its binary form [1]. Our encryption technique processes directly on symbols without passing to the bit level. In addition, this technique does not need any hardware; it is totally based on software. This technique is much simpler than existing techniques, thus a more robust and efficient system is achieved. The following sections discuss the proposed scheme: Section 2 enumerates the security requirements of mobile networks. Section 3 gives a quick overview of existing GSM encryption algorithms and a variety of attacks on these algorithms. Section 4 illustrates the proposed End-To-End encryption method. Section 5 the simulation results, and finally, concludes this paper by summarizing the key points and proposing related suggestions.

## 2. SECURITY REQUIREMENTS OF MOBILE NETWORKS

Security has become an essential topic in current mobile and wireless networks. As the security procedures for such networks elevates, the tools and techniques used to attack such networks also increases. Wireless communications security is the measures or methods used to protect the communication between certain entities. To protect the entity from any third party attacks, such as revealing a particular identity, data modification or data-hijacking, eavesdropping, impersonating an identity, Protection mechanisms are used. Devoted technologies for securing data and communication are mandatory in wireless networks, which vary according to the category of wireless technology deployed. Security in mobile networks handles a diversity of issues, from authenticating a user accessing a network, to data integrity and data encryption. GSM, like a lot of other systems with huge users' numbers, contains numerous precious resources that need protection against misuse and deliberate attacks. This section highlights the GSM Network precious resources, which are important to protect for the best of the system's shareholders.

The facilities listed below are provided to insure security to the users of the communication networks [3]-[7]:

**Confidentiality**: This means that the transmitted information is only disclosed to the authorized parties. Sensitive information disclosed to an adversary could have severe consequences.

**Integrity**: This assumes that a message is not altered in transit between sender and receiver. Messages could be corrupted due to network malfunctioning or malicious attacks.

**Authentication**: Authentication guarantees the identity of the entity with which communications are established, before granting it the access to the resources of the network. In the absence of authentication mechanisms, an attacker could masquerade as a legitimate entity and attempt to violate the security of the network.

**Nonrepudiation**: This means that the source of a message cannot deny having sent the message. An attacker could generate a wrong message that appears to be initiated from an authorized party, with the aim of making that party the guilty one. If non-repudiation is guaranteed, the receiver of a wrong message can prove that the originator has transmitted it, and that, therefore, the originator misbehaved.

**Access control**: Access control means that only authorized parties can be allowed to access a service on the network, use a resource, or participate in the communications; any other entity is denied access. The access control assumes the authentication of the entity trying to get access to the network.

**Network availability**: Availability ensures that all resources of the communications network are always utilizable by authorized parties. An attacker may launch a Denial of Service (DoS) attack by saturating the medium, jamming the communications, or keeping the system resources busy in any other way or by any other means. The aim here is just to slow down or stop authorized parties from having access to the resources, thereby making the network unusable.

## 3. GSM encryption and attacks

In GSM, A5 stream cipher is used [5]. Versions A5/1 and A5/2 were kept secret for a long period of time. Briceno et al reverse-engineered A5/1 and A5/2 from a GSM handset and published them. After which, attacks were rapidly found for these algorithms. The principal problem is the small key length of the session key Kc. The actual length of Kc is 64 bits.

However, only 54 bits are effective. Even though this key size is sufficiently big to protect against real-time attacks, the hardware state available today makes it possible to record the packets between the mobile subscriber and the BTS and then decrypt them afterward [6].

Biryukov et al. found a known-key stream attack on A5/1 that needed about two seconds of the key stream and recovers Kc in a few minutes on a PC after a large pre-processing stage. Barkan et al. [5] have proposed a ciphertext-only attack on A5/1 that also recovers Kc using only four frames; the problem was its complexity. A5/2 was also cracked and proved to be totally vulnerable. The attack needed very few pseudo random hits and only 216 steps [5].

A new security algorithm, known as A5/3 provides users of GSM mobile phones with an even higher level of protection against eavesdropping than they have already. A5/3 is based on the Kasumi algorithm, specified by 3GPP for use in 3G mobile systems. The A5/3 encryption algorithm particularly provides signaling protection to protect important information such as telephone numbers as well as user data protection to secure voice calls and other user generated data. This algorithm were so far assumed to be stronger than A5/1 and A5/2, but the Biham et al attack shows that the key can be obtained quickly without applying exhaustive key search.

## 4. END-TO-END encryption method

### 4.1. Review of GSM Voice Transmission

The process of GSM voice channel transmission illustrated in Figure 1, includes five components: A/D module, RPE-LTP vocoder, channel coding/decoding module, wireless encryption /decryption module, and GMSK modulation/demodulation module. The wireless encryption/decryption module only works on wireless channel. So it cannot provide end-to-end secure communication in GSM system.

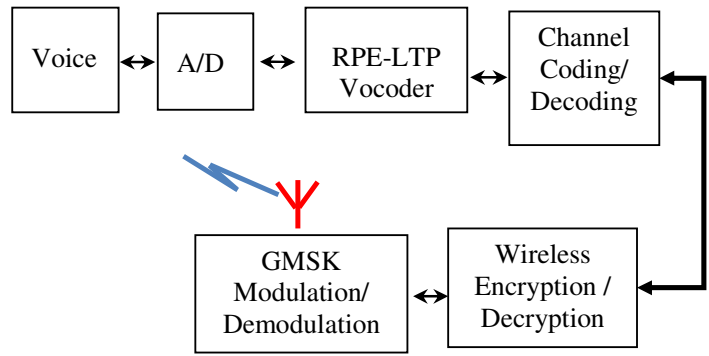

Figure 1. GSM voice transmission process.

### 4.2. Review of GSM Voice Transmission

The RPE-LTP [4]-[2] is an important algorithm in the field of voice encoding. It is not only used in GSM, but also used in Internet.

At the transmitter, the processing in the RPE-LTP Encoder includes pre-processing, LPC analysis, short-term analysis filtering, Long-Term Prediction and Regular Pulse Excitation sequence coding. The details is described as following: first Encoder samples original digital voice signal at 8kHz sampling rate, and removes the direct current component, then it can make use of FIR filter to pre-emphasis the high frequency. Secondly, LPC analysis takes every 160 sample points (20ms) as one frame and figures out 8 logarithm acreage ratio parameter for each frame. Short-time analysis filter produces LPC residual signal. It removes redundancy farther coding with RPE-LTP, and outputs 260 bits coding every frame at last. At the receiver, it practices a reverse processing and rebuilds the original speech signal.

### 4.3. Voice Encryption Method

In a general way, encryption/decryption module is put before the RPE-LTP vocoder, which is easy to implement in MT (Mobile Terminal). But it cannot accomplish the end-to-end secured communication, and need to be modified in BS (Base Station). So a novel voice encryption/decryption method is proposed based on voice channel, which can fulfill the end-to-end secured communication without any modification in BS. The novel voice encryption/decryption scheme is depicted in Figure 2.

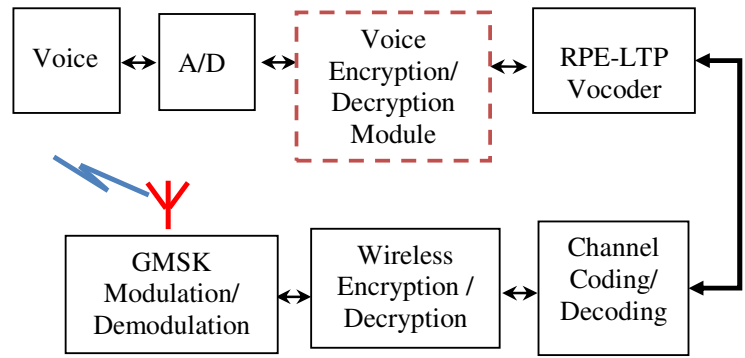

Figure 2.Voice encryption module access point in MT.

In Figure 2, the new voice encryption/decryption module is inserted between A/D and RPE-LTP vocoder in MT. The coming voice signal from the A/D module would firstly arrive to the newly-added Voice Encryption/Decryption module and finishes the encryption. After that, it is sent to the RPE-LTP vocoder. Hence, this encryption method must penetrate the RPE-LTP vocoder and have ideal encryption intensity. Simultaneously this encrypted signal can be recovered to get the original understandable speech at the receiver. This new voice encryption method is a kind of signal source encryption technology, so it could achieve the end-to-end secured communication.

### 4.4. Encryption algorithm

Principle of the encryption algorithm:

For implementing encryption algorithm, we follow the following steps:

- Decomposition of a speech signal in sub-frame, each frame is represented by an index.

- Encrypting data with inversion and random permutation algorithm, which gives the permutation indexes.

- Encrypting these indexes with DES algorithm.

- Used these indexes to decrypt the signal.

In this paper, we propose an algorithm that combines between permutation and inversion of the voice signal samples, giving as a result the permuted indexes. These indexes are processed in an encryption/decryption module by DES algorithm, and finally, these encrypted permuted indexes are added to the compressed encrypted voice signal samples after the RPE-LTP module. So it has a good recovery character to RPE-LTP vocoder, and its encryption intensity also can meet the special requirement. The algorithm is mainly intended to make the encrypted voice signal to be similar to the natural human voice signal, and can penetrate the RPE-LTP vocoder, and then it can execute all the encryption and decryption process. (See Figure 3).

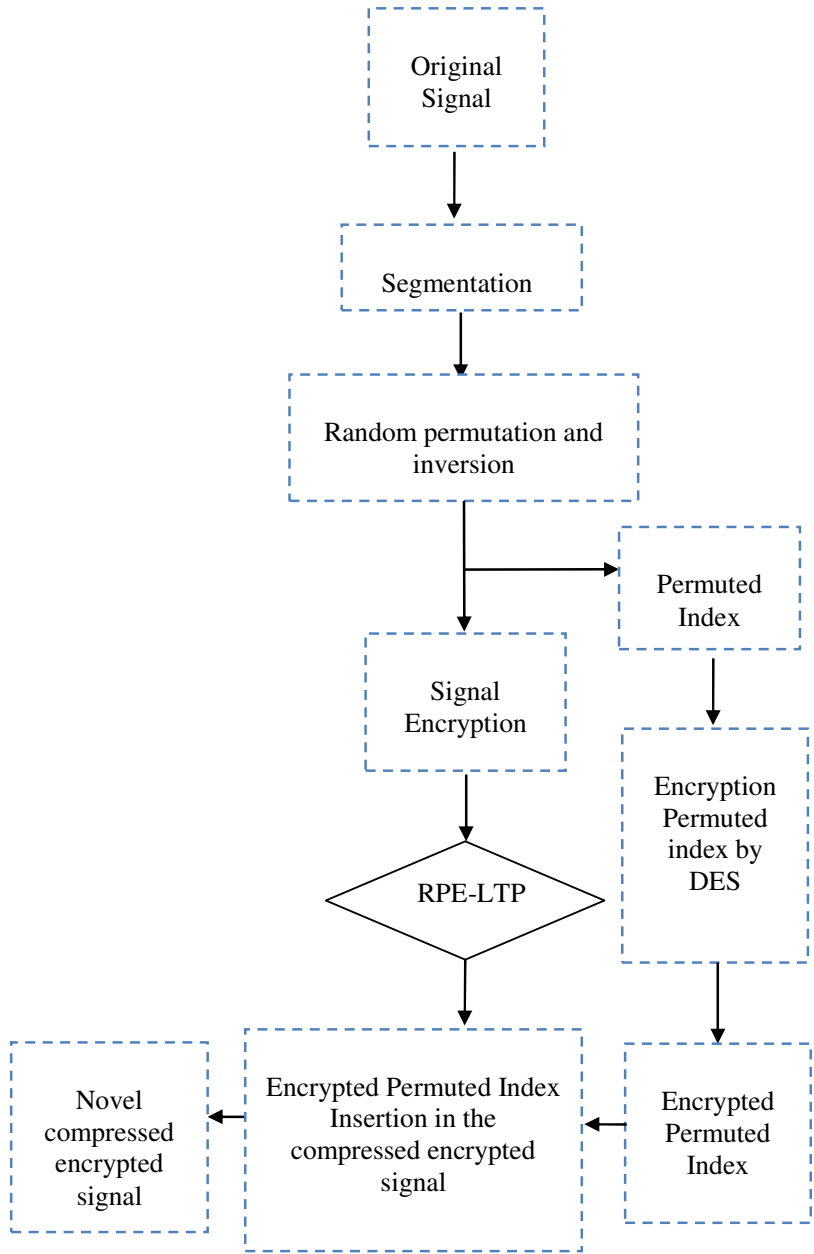

Figure 3. Encryption chart

## 5. Simulation results

This section presents the results for the proposed method adopted in Section 3. This section also discusses the obtained results from implementing the system. In order to implement such a system, one must go through several steps which were described in details in the preceding sections. The implementation for this simulated project is written by MATLAB.

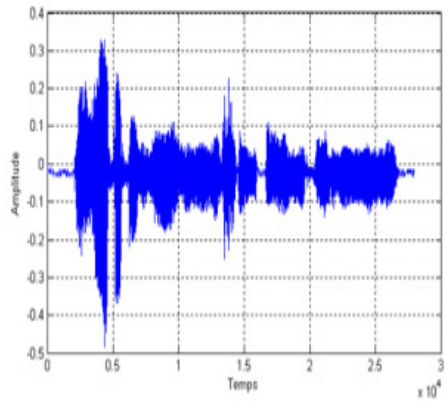
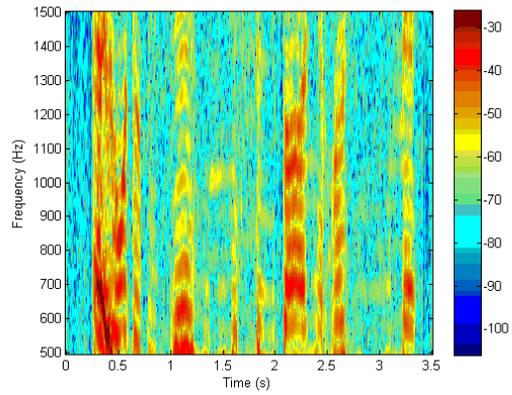

a.1 Original temporal signal

b.1 Spectrogram of original Signal

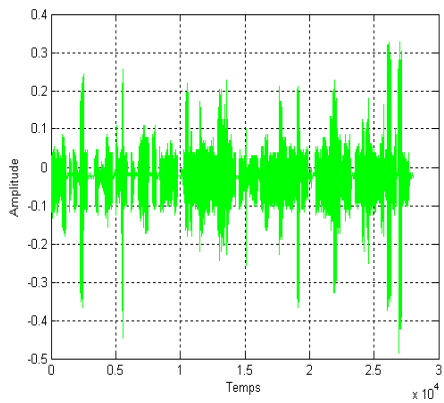
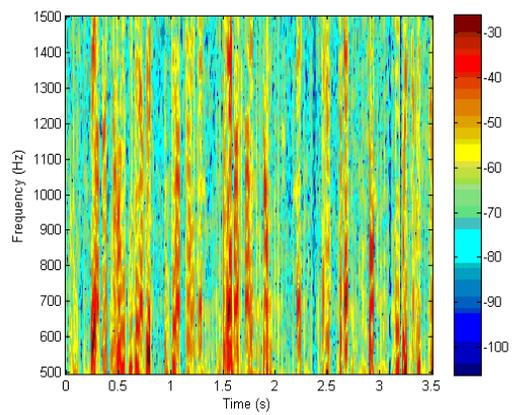

a.2 Encrypted temporal signal before LPC

b.2 Spectrogram of encrypted signal before LPC

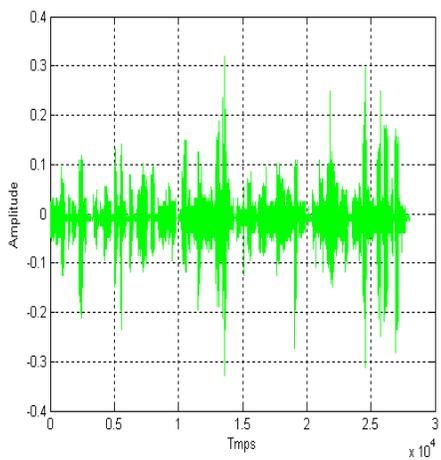
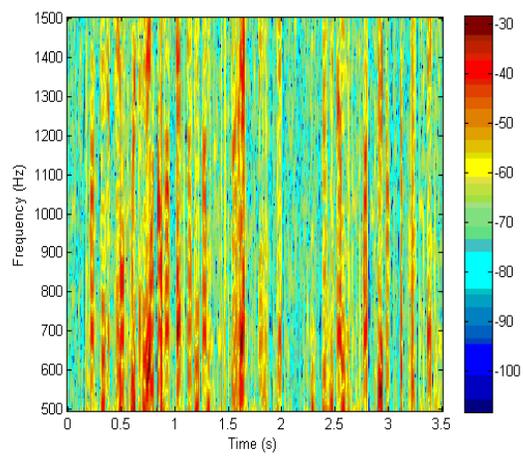

a.3 Synthesized encrypted temporal signal.

b.3 Spectrogram of Synthesized encrypted Signal.

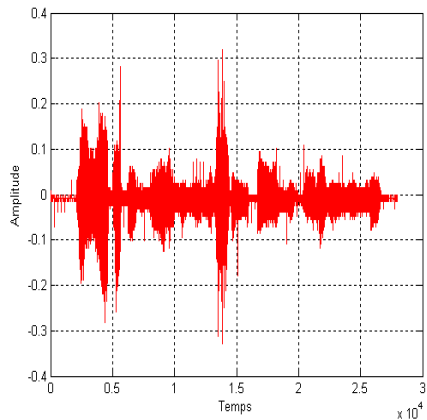
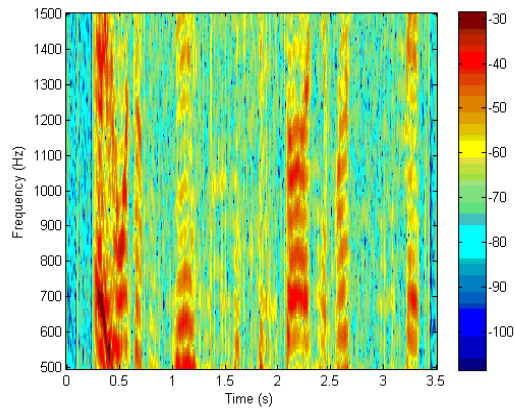

a.4 Synthesized deciphered temporal signal.

b.4 Spectrogram of Synthesized deciphered Signal.

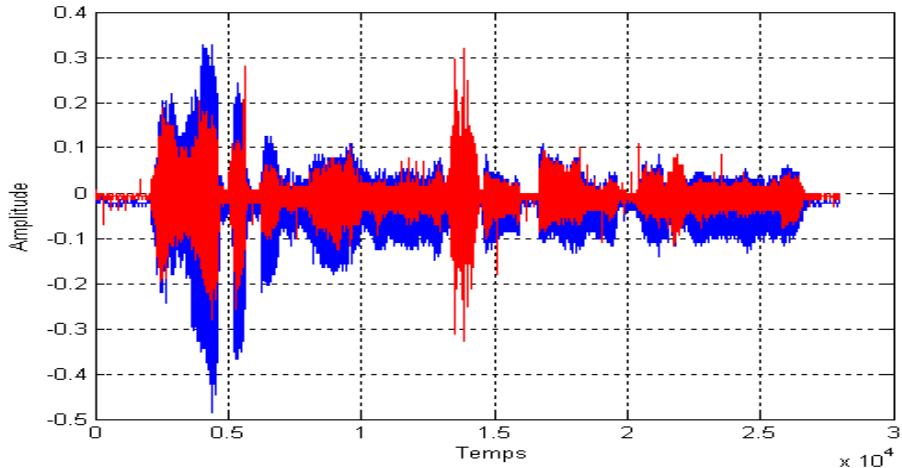

C. Comparison between the original signal and the synthesized signal.

In figure C, the margin between the original signal (blue) and the synthesized one (red) is due to the reduction of the bit rate imposed by the RPE-LTP module.

## 6. Conclusion

In this paper, a novel kind of encryption method is proposed to fulfill the end-to-end secured communication in the GSM voice channel. The new encryption method solves the problem that traditional encryption algorithms cannot be used in voice channel directly because of RPE-LTP vocoder requirements in GSM system. In addition, this encryption method has the advantages of suiting the RPE-LTP compression module requirements, good compatibility to GSM networks, and suitable implementation without any adjustment in current GSM signalling system. The algorithm presented in this paper is made by the DES algorithm, but it can be done also by other encryption methods such as: RSA, RC4 and AES.

## REFERENCES

[1]     David G. W. Birch and Ian J. Shaw, "Mobile communications security private or public", IEEE, June 1994.

[2]     ETSI Speech processing functions General Description, GSM06.01-1999, Version 8.0.1 pp22-53.

[3]     H.Imai, M.G. Rahman, K. Kobara "Wireless Communications Security" ARTECH HOUSE 2006.

[4]     K.Hellwig, P.Vary, D. Massaloux, ectl. Speech Codec for the European Mobile Radio System. IEEE Global Commu Conf, 1989: pplO65-1069.

[5]      Ross Anderson, Mike Roe "A5 - The GSM Encryption Algorithm", 1994.

[6]      Dr. S. Muhammad Siddique and Muhammad Amir "GSM Security Issues and Challenges" (SNPD'06) 2006.

[7]      Noureddine Boudriga "Wireless Communications Security",CRC 2010 by Taylor and Francis Group, LLC.

**Authors**

Khaled Merit received the Engineer degree in Telecommunication from National Institute of Telecommunication & ICT Oran June 2007, and magister degree in Telecommunication from National Institute of Telecommunication & ICT Oran January 2011, currently he is PhD Student at the University Tlemcen in the area of Security and Cryptography.

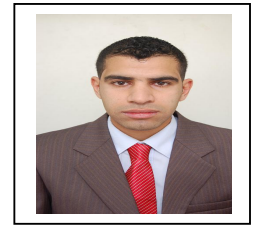